\documentclass[journal,twoside,web]{ieeecolor}
\usepackage{tdei}
\usepackage{amsmath,amsfonts}
\usepackage{array}
\usepackage{multirow}
\usepackage[caption=false,font=normalsize,labelfont=sf,textfont=sf]{subfig}
\usepackage{textcomp}
\usepackage{stfloats}
\usepackage{url}
\usepackage{graphicx}
\usepackage{siunitx}
\usepackage{cite}
\usepackage[capitalize]{cleveref}
\crefformat{equation}{(#2#1#3)}
\Crefformat{equation}{(#2#1#3)}

\makeatletter
\renewcommand{\fnum@figure}{\textcolor{subsectioncolor}{\sf Fig.~\thefigure}}
\renewcommand{\fnum@table}{\textcolor{subsectioncolor}{\sf TABLE~\thetable}}
\makeatother

\begin{document}

\title{Electric Field Optimization of High-Voltage Vacuum Feedthroughs}

\author{Lin Si, Evan Angelico, and Giorgio Gratta
\thanks{Manuscript received XX Month 2026; revised XX Month 2026; accepted XX Month 2026.  \textit{(Corresponding author: Lin Si.)}}
\thanks{The authors are with the Department of Physics, Stanford University, Stanford, CA 94305 USA (e-mail: linsi@stanford.edu).}
\thanks{E. Angelico is now with TAE Technologies, Foothill Ranch, CA USA.}}

\markboth{IEEE Transactions on Dielectrics and Electrical Insulation,~Vol.~XX, No.~X, Month~2026}%
{Si \MakeLowercase{\textit{et al.}}: Electric Field Optimization of High-Voltage Vacuum Feedthroughs}

\IEEEpubid{0000--0000/00\$00.00~\copyright~2026 IEEE}

\maketitle

\begin{abstract}
High-voltage vacuum feedthroughs based on brazed alumina ceramic are designed for hermeticity, low outgassing, and integrity of the metal--ceramic joint, while the radius of the center conductor receives comparatively little attention. The optimization of such feedthroughs, aimed at minimizing the electric field on the vacuum side of the device, is reported. Commercial feedthroughs are found to have center conductor diameters that are too small, resulting in unnecessarily large fields. The problem is studied analytically, treating the feedthrough as a layered-dielectric-conductor coaxial system, and the result is refined with an axisymmetric finite element model of two commercial devices. For a 100 kV feedthrough, the peak surface field is minimized when the center conductor radius is increased from 2 mm to about 7 mm, which lowers that field by 27\% in vacuum and by 30\% in liquid xenon, while a 30 kV feedthrough yields smaller gains of a few percent. The optimum radius depends on the permittivity of the medium that fills the vacuum side. A simple retrofit is also presented: a stainless steel sleeve with hemispherical ends and internal flutes that preserve pumping conductance, which realizes the optimum radius without compromising the outgassing properties of the device. This work is important for applications whereby the ``vacuum side'' of the feedthrough is, in fact, filled with a dielectric which may not have the voltage rigidity of vacuum.
\end{abstract}


\section{Introduction}
\IEEEPARstart{H}{igh-voltage} (HV) vacuum feedthroughs are commercially produced and are designed for hermeticity and integrity of the brazed ceramic--metal joint, low outgassing, and the ability to operate over a broad range of temperatures~\cite{Schwartz:2003, Weterings:2002}. The HV design is generally limited to guaranteeing proper performance on the air side, and to the metal--ceramic--vacuum triple junction, which is a region known to be susceptible to flashover and joint failure~\cite{Weterings:2002}. The radius of the center conductor itself has received comparatively little attention. This is in contrast with HV coax cables, where the center conductor radius is optimized by taking into account the electric field on its surface.

In recent times, HV feedthroughs based on brazed alumina ceramic have also been used for applications, such as radiation detectors, in which the ``vacuum side'' of the feedthrough is, in fact, filled with a dielectric which may not have the same dielectric strength as high vacuum. This is the case for liquid Xenon (LXe)~\cite{Aprile:2017, Akerib:2020, PandaX}, and liquid Argon (LAr) Time Projection Chambers (TPCs)~\cite{Acciarri:2017, Amerio:2004, Abi:2020}. The use of other liquefied noble gases has also been investigated for calorimetry in high energy physics~\cite{Peleganchuk:2009} and the detection of rare events~\cite{Hertel:2019, Bandler:1992}. For these applications, ultra-high vacuum performance (ideally $10^{-9}~\mathrm{mbar\,L/s}$ or better) is a useful proxy for low outgassing. It indicates that the liquid will not be contaminated by electronegative impurities, which would otherwise spoil the long electron lifetimes the detectors require. At the same time, these detectors must resolve signals of only a few electrons liberated in the liquid. Any spurious emission of electrons, even far below breakdown, is therefore intolerable, and all HV components must be carefully optimized. Since the field-emission current depends exponentially on the local surface field~\cite{Latham:1995}, reducing the peak field on the cathode surfaces, including the central conductor of the feedthrough, is desirable.
\IEEEpubidadjcol

\begin{figure}[!t]
\centering
\includegraphics[width=3.4in]{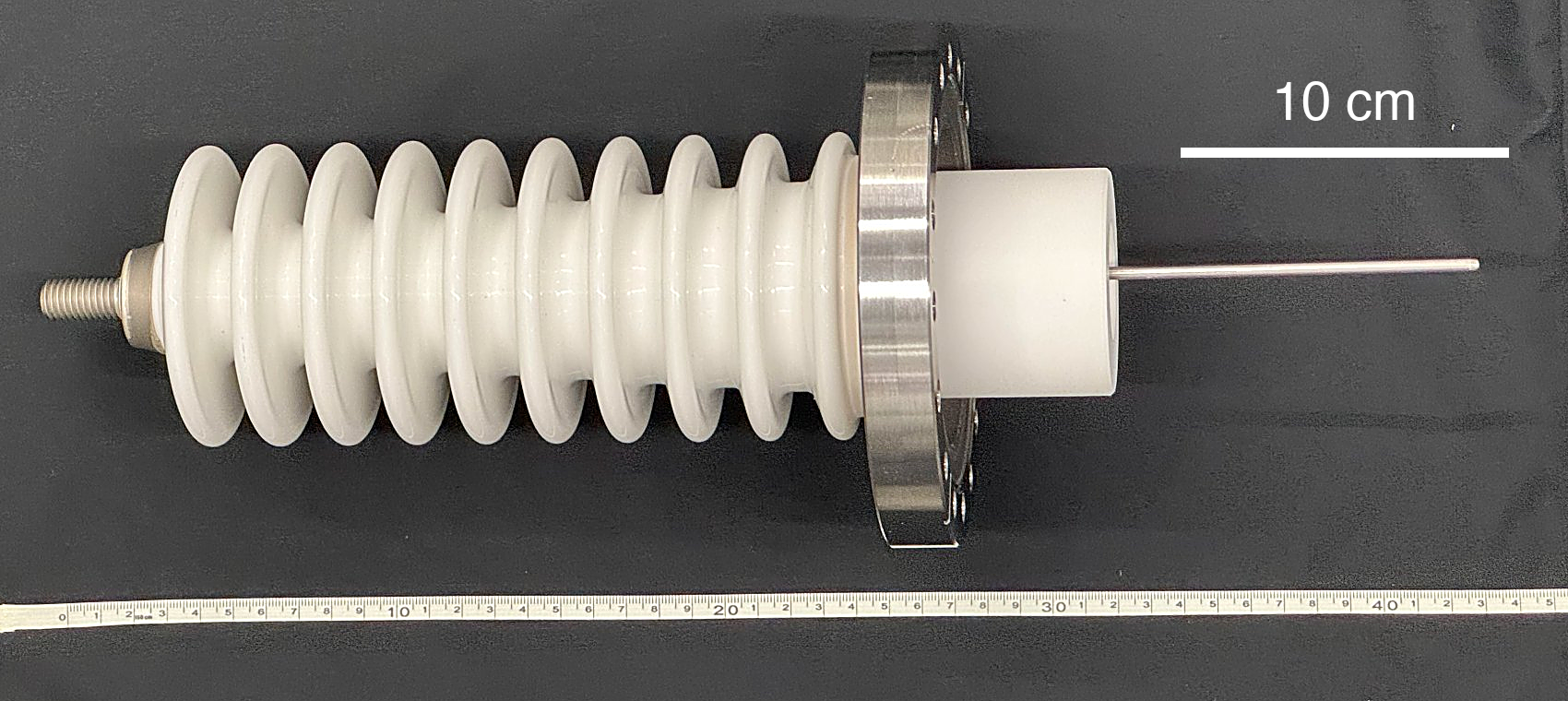}
\caption{The commercial \SI{100}{\kilo\volt} feedthrough studied in this work, with the vacuum side on the right and the air side on the left.}
\label{fig:feedthrough}
\end{figure}

This study shows that the highest field in a commercial \SI{100}{\kilo\volt} CF-flange feedthrough (Kurt J.\ Lesker model EFT1C12156A, \cref{fig:feedthrough}) can be decreased by $\simeq 30\%$ by tripling the radius of the center conductor. The feedthrough consists of a stainless steel center conductor, a brazed alumina insulator, and a DN100CF flange; the alumina is corrugated on the air side and smooth on the vacuum side. The final optimization depends on the relative permittivity of the medium on the ``vacuum'' side.

In practice, the radius resulting from the optimization can be obtained by installing a metallic sleeve over the center conductor. In the solution presented here, a form of rifling on the internal surface of the sleeve preserves the pumping conductance needed for vacuum operation.

The paper is organized as follows: the optimum radius is first derived analytically for a layered-dielectric-conductor coaxial arrangement; this approximation is then improved, for specific cases, using a finite-element model. Finally, a practical sleeve design and the implications for medium-dependent optimization of commercial HV feedthroughs are discussed.

\section{Electric Field Optimization}
The physical origin of the optimization is well known. First, consider the simplified limit in which the entire region between the center conductor and the grounded shell is filled by a single dielectric. The surface field on the center conductor is
\begin{equation}
\label{eq:Esingle}
E(a) = \frac{V}{a\,\ln(c/a)},
\end{equation}
where $V$ is the applied potential, and $a$ ($c$) is the radius of the center conductor (grounded shield). $E(a)$ diverges both as $a\to 0$ (the $1/a$ prefactor) and as $a\to c$ (the logarithm vanishes); an interior minimum is reached at $a^{\star} = c/e$, where $e$ is the base of natural logarithms~\cite{Jackson:1999}.

Although the actual geometry of feedthroughs is more complex, it is instructive to first simplify the problem to that of layered-dielectric coaxial cylinders, so that an analytical solution exists. Schematically:
\begin{itemize}
    \item a center conductor (at HV) of radius $a$, which is treated as a free parameter;
    \item a dielectric gap in the radial region, $a \le r \le b$, of relative permittivity $\varepsilon_1$; two cases are considered, $\varepsilon_1 = 1$ (vacuum) and $\varepsilon_1 = 1.85$ (LXe~\cite{Xu:2019});
    \item an alumina insulator, for $b \le r \le c$, of relative permittivity $\varepsilon_2 = 9.8$ (values in the range $9.4$--$10.6$ are reported for sintered alumina~\cite{Boumous:2025});
    \item an outer cylinder at $r = c$, which provides the ground.
\end{itemize}
The values of $b$ and $c$ are assumed to be fixed while the center conductor radius, $a$, is varied in the optimization, displacing variable amounts of the internal medium of permittivity $\varepsilon_1$.

The field at the surface of the center conductor is
\begin{equation}
\label{eq:Emax}
    E_1(a) \;=\;
    \frac{V}{a\bigl[\,\ln(b/a) + (\varepsilon_1/\varepsilon_2)\ln(c/b)\,\bigr]},
\end{equation}
which is minimized when $a$ assumes the value
\begin{equation}
\label{eq:astar}
    a^{\star} \;=\; \frac{b}{e}\,
    \Bigl(\frac{c}{b}\Bigr)^{\varepsilon_1/\varepsilon_2}.
\end{equation}

For the \SI{100}{\kilo\volt} feedthrough, $b = \SI{16}{\milli\metre}$ and $c = \SI{36}{\milli\metre}$ are the inner radius of the alumina and the inner radius of the grounded flange in the region of the peak field, taken from the manufacturer's CAD model. With these values, \cref{eq:astar} predicts an optimum center conductor radius of $a^{\star} = \SI{6.4}{\milli\metre}$ in vacuum and $a^{\star} = \SI{6.9}{\milli\metre}$ in LXe, in both cases substantially larger than the original radius $a_0 = \SI{2}{\milli\metre}$. Substituting \cref{eq:astar} into \cref{eq:Emax}, the bracket in the denominator equals unity, so that $E_1(a^{\star}) = V/a^{\star}$, and the peak-field reduction relative to the original conductor is
\begin{equation}
\label{eq:reduction}
    1 - \frac{E_1(a^{\star})}{E_1(a_0)}
    = 1 - \frac{a_0}{a^{\star}}\Bigl[\ln\frac{b}{a_0} + \frac{\varepsilon_1}{\varepsilon_2}\ln\frac{c}{b}\Bigr]
    =
    \begin{cases}
    32.4\%, & \text{vacuum},\\
    34.9\%, & \text{LXe}.
    \end{cases}
\end{equation}
Varying $\varepsilon_2$ within the reported alumina range results in negligible changes in the maximum field obtained.

\subsection{Finite-Element Model}
A finite element (FE) model can further incorporate the specific features of the system, which deviate from the infinite coaxial cylinder geometry. The electrostatic (Laplace) problem was solved with the open-source \textsc{Elmer} FE package~\cite{Malinen:2013}.
A two-dimensional axisymmetric mesh was generated in \textsc{gmsh}~\cite{Geuzaine:2009}. A distance-based background mesh-size field was applied. The refinement features were the center conductor surface, the alumina inner surface, and the metal--ceramic--vacuum (or LXe) triple junctions. Elements within \SI{0.3}{\milli\metre} of these features had a size of \SI{0.1}{\milli\metre}, increasing linearly to \SI{3}{\milli\metre} at a distance of \SI{5}{\milli\metre}. The center conductor was held at \SI{-1}{\kilo\volt} and all metallic surfaces of the flange at ground; field magnitudes quoted below are at this applied voltage and scale linearly to any operating voltage of interest.

As a check on the numerical accuracy of the solutions, mesh convergence was verified for both the as-built ($r=\SI{2.0}{\milli\metre}$) and optimum-radius ($r=\SI{7.0}{\milli\metre}$) configurations in vacuum by halving the local element size near the refinement features from \SI{0.1}{\milli\metre} to \SI{0.05}{\milli\metre}; the peak surface field changed by less than \SI{1}{\percent} in both cases.

\begin{figure*}[!t]
\centering
\subfloat[]{\includegraphics[width=3.4in]{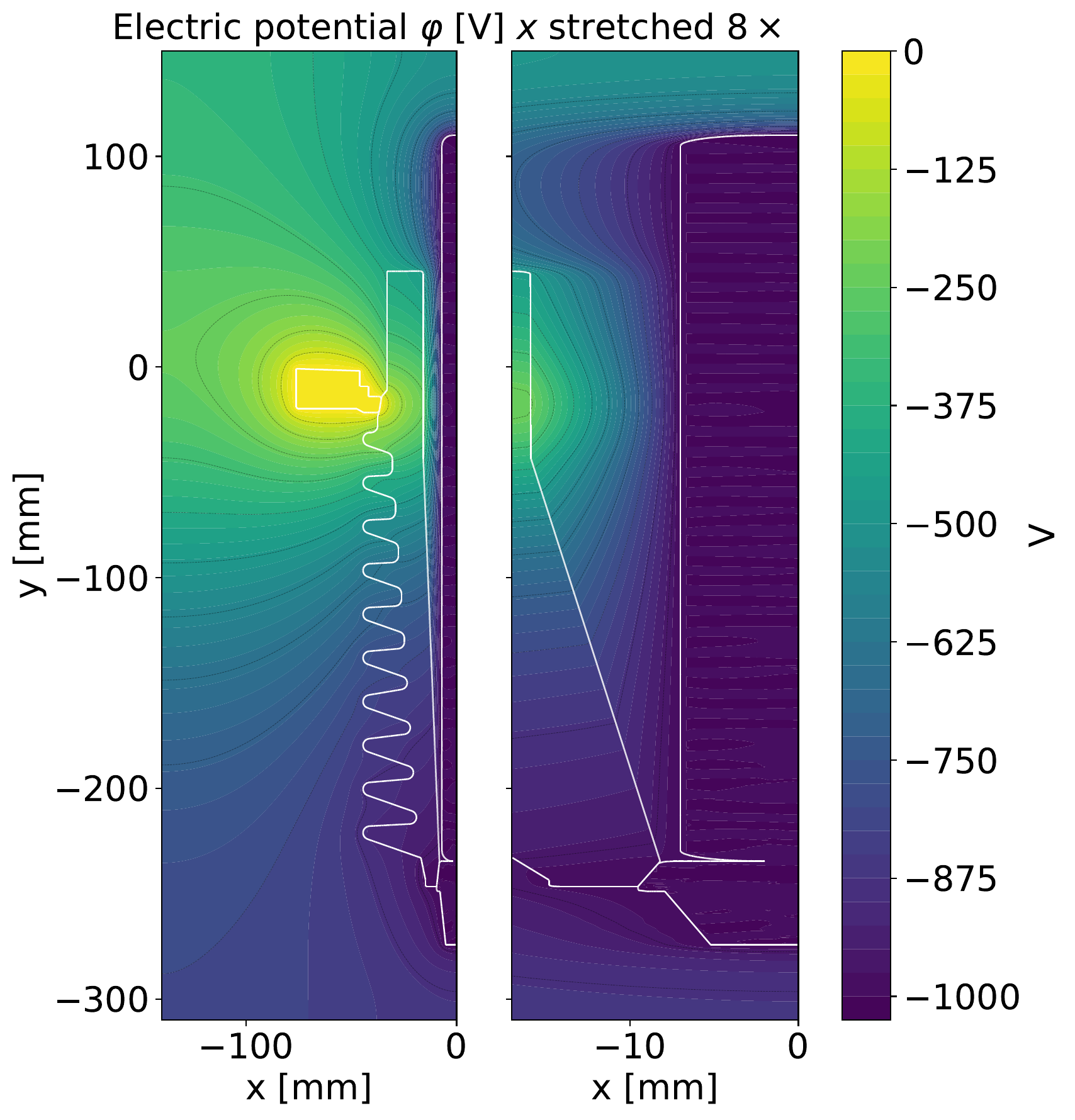}%
\label{fig:potential}}
\hfil
\subfloat[]{\includegraphics[width=3.4in]{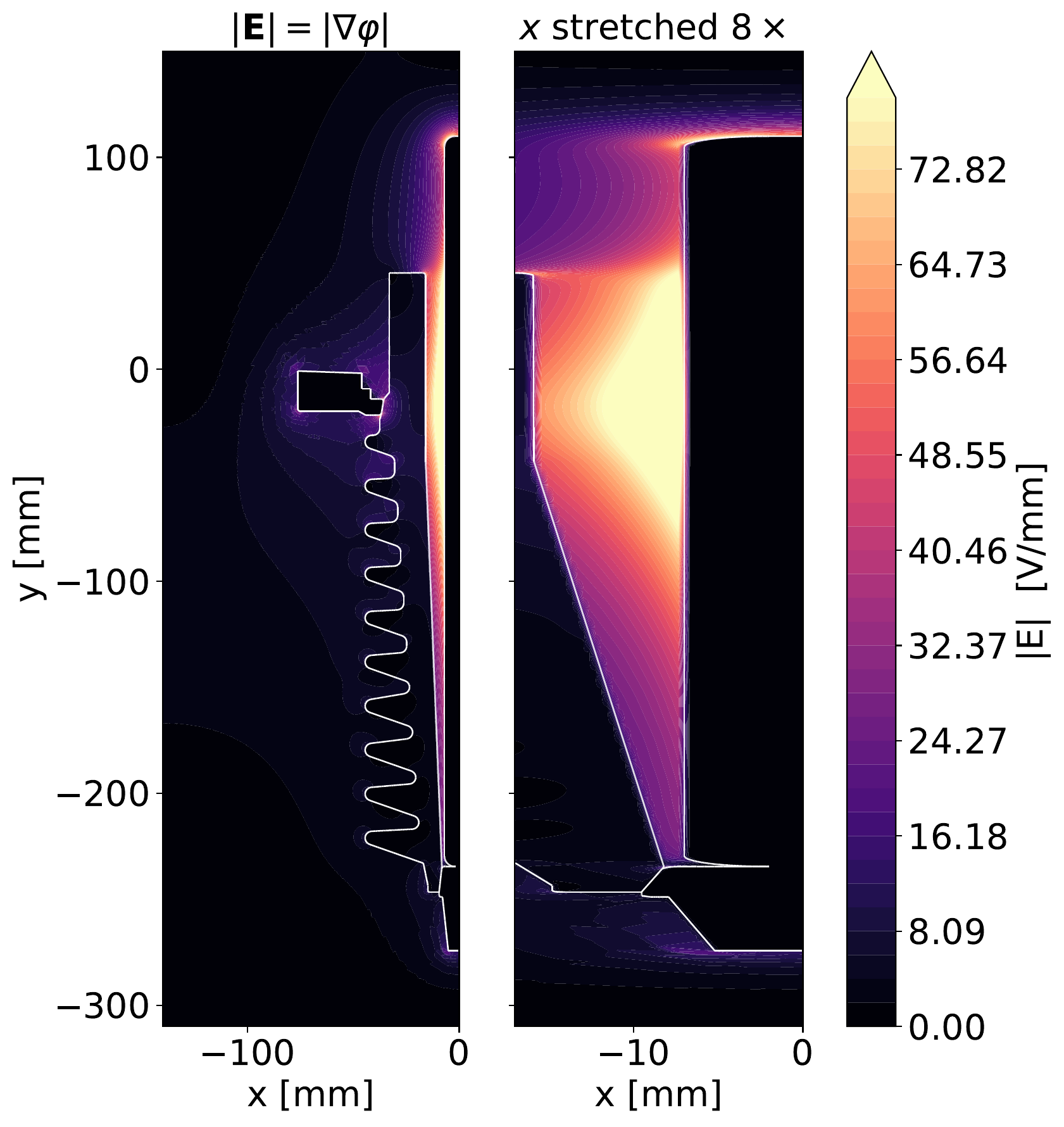}%
\label{fig:fieldmag}}
\caption{Finite-element solution for the commercial \SI{100}{\kilo\volt} feedthrough in vacuum with the sleeve ($r = \SI{7}{\milli\metre}$) increasing the radius of the inner conductor. The center conductor is held at \SI{-1}{\kilo\volt}. (a)~Electric potential $\varphi$. (b)~Field magnitude $|\mathbf{E}|=|\nabla\varphi|$. In each subfigure the left view shows the full geometry at true ($1\!:\!1$) scale, while the right view is an anamorphic close-up of the near-conductor region $x\in[-17,0]\,\si{\milli\metre}$ in which the radial ($x$) axis is stretched by a factor of $8$ relative to the axial ($y$) axis. In this view the near-conductor structure becomes more clearly visible. The global maximum of $|\mathbf{E}|$ sits at the surface of the center conductor near the flange transition.}
\label{fig:fieldmaps}
\end{figure*}

Representative FE solutions for the \SI{100}{\kilo\volt} feedthrough operated in vacuum, with the center conductor enlarged to the optimum radius $r = \SI{7}{\milli\metre}$ by the sleeve of Section~III, are shown in \cref{fig:fieldmaps}. The electrostatic potential (\cref{fig:fieldmaps}(a)) drops smoothly from the center conductor (at \SI{-1}{\kilo\volt}) to the grounded flange, mostly across the vacuum gap between the conductor and the alumina, owing to the higher permittivity of the alumina, consistent with the layered-dielectric setup of the analytical model. The field magnitude (\cref{fig:fieldmaps}(b)) is strongly localized on the center conductor surface, particularly near the flange transition; because this localization is otherwise squeezed onto a near-vertical line, it is shown most clearly in the anamorphic close-up (right), where the radial direction is stretched.

\begin{figure*}[!t]
\centering
\includegraphics[width=5.6in]{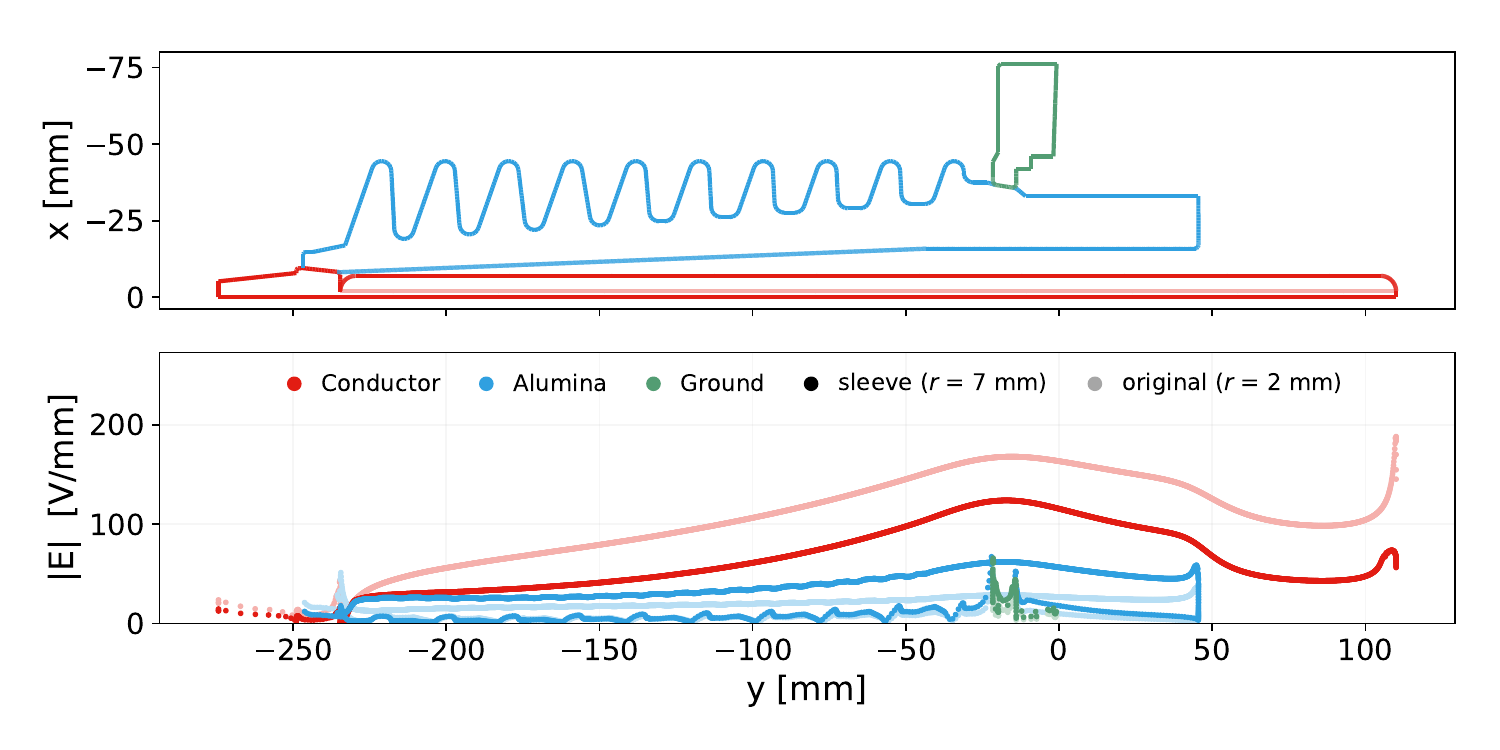}
\caption{Field along the boundaries of the \SI{100}{\kilo\volt} feedthrough in vacuum, with the center conductor at \SI{-1}{\kilo\volt}. Top: geometry, showing the center conductor (red), alumina (blue), and grounded flange (green); the original conductor ($r = \SI{2}{\milli\metre}$) is shown in light red and the sleeve ($r = \SI{7}{\milli\metre}$) in red. Bottom: $|\mathbf{E}|$ sampled \SI{0.5}{\milli\metre} from each boundary as a function of the axial coordinate $y$, with the sleeve (solid colors) and with the original conductor (light colors). The spike of the original conductor at $y \simeq \SI{110}{\milli\metre}$ is due to the flat end of the modeled pin.}
\label{fig:boundary}
\end{figure*}

To quantify the field on the three relevant boundaries, $|\mathbf{E}|$ was extracted \SI{0.5}{\milli\metre} from the center conductor, the alumina insulator, and the grounded flange as a function of the axial coordinate $y$ (\cref{fig:boundary}), for the original conductor and for the sleeve of Section~III ($r = \SI{7}{\milli\metre}$). The field is sampled at this distance because it is singular at the triple junctions. For the original conductor, the field on the conductor reaches \SI{169.2}{\volt\per\milli\metre} near $y \simeq \SI{-20}{\milli\metre}$, where the alumina inner radius first opens up, while the fields on the alumina and on the flange remain lower by a factor of ${\sim}4$, confirming that the layered-coax approximation captures the dominant physics of the active region. With the sleeve, the peak on the conductor drops to \SI{123.9}{\volt\per\milli\metre}, and the field at the bottom of the alumina bore, where the conductor meets the alumina (cathode triple junction), is reduced by about a factor of two. Near the flange, the fields on the alumina and at the flange triple junction increase from about 40 to \SI{67}{\volt\per\milli\metre}, but remain about half of the peak on the conductor; since the flange is the anode for the negative polarity considered here, this increase is not expected to enhance electron emission. The same pattern is found in LXe.

\subsection{Optimum Center Conductor Radius}
The center conductor radius, $r$, was then scanned from the as-built value to the maximum value compatible with the inner bore of the alumina cylinder, separately for $\varepsilon_1$ corresponding to the vacuum and LXe cases. A different, commercial \SI{30}{\kilo\volt} feedthrough (Kurt J.\ Lesker model EFT3012093) was also studied with the FE solver. The radial step size was \SI{0.5}{\milli\metre} for the \SI{100}{\kilo\volt} feedthrough (spanning $r = \SI{2}{\milli\metre}$ to \SI{8}{\milli\metre}) and \SI{0.2}{\milli\metre} for the \SI{30}{\kilo\volt} feedthrough ($r = \SI{1.2}{\milli\metre}$ to \SI{3.4}{\milli\metre}). Radii larger than $r_0$ were modeled as a sleeve over the original pin, extending over the whole length of the conductor and terminated at both ends by rounded caps of radius $r - r_0$, as in the prototype of Section~III.

\begin{figure*}[!t]
\centering
\subfloat[]{\includegraphics[width=3.4in]{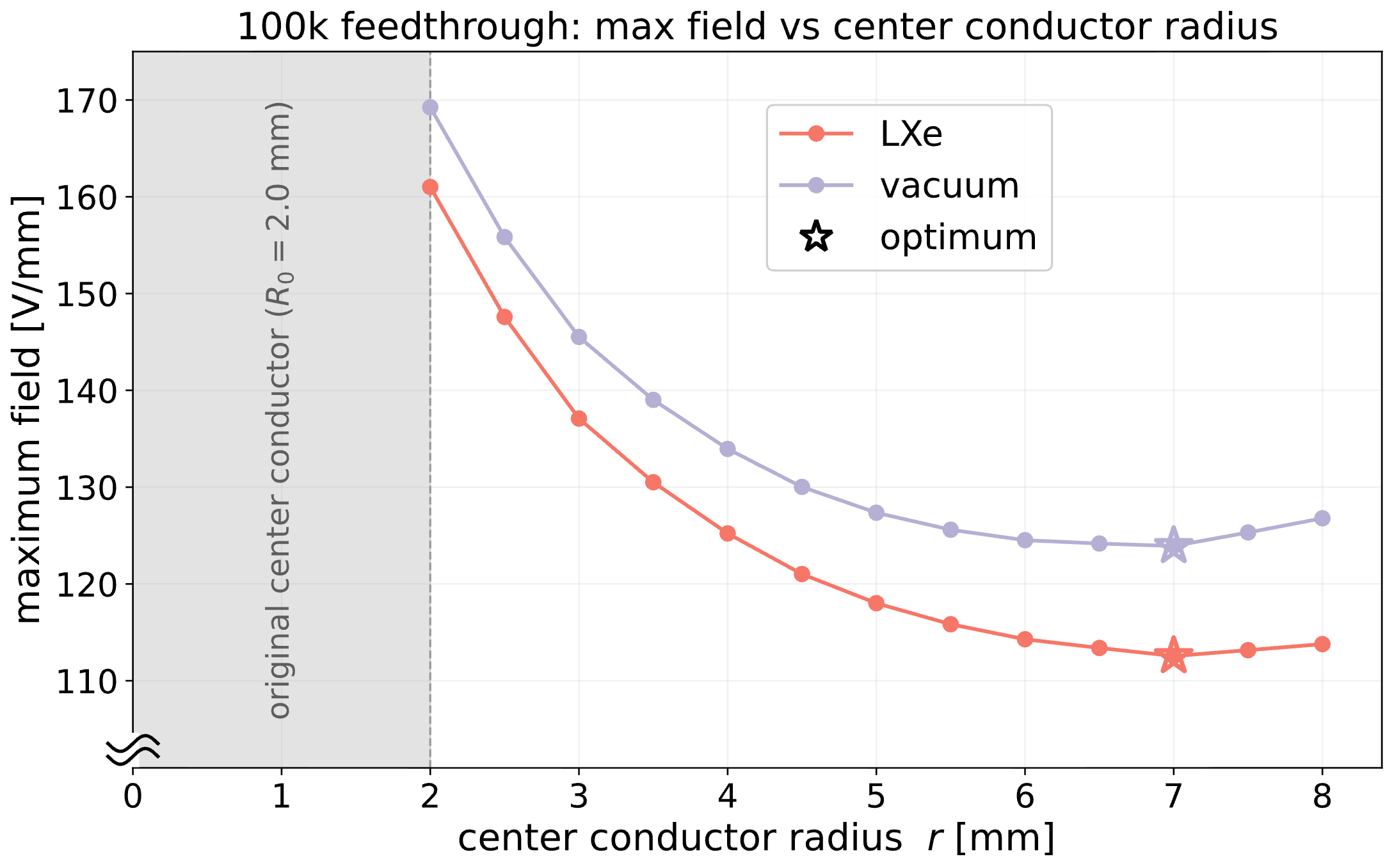}%
\label{fig:scan100k}}
\hfil
\subfloat[]{\includegraphics[width=3.4in]{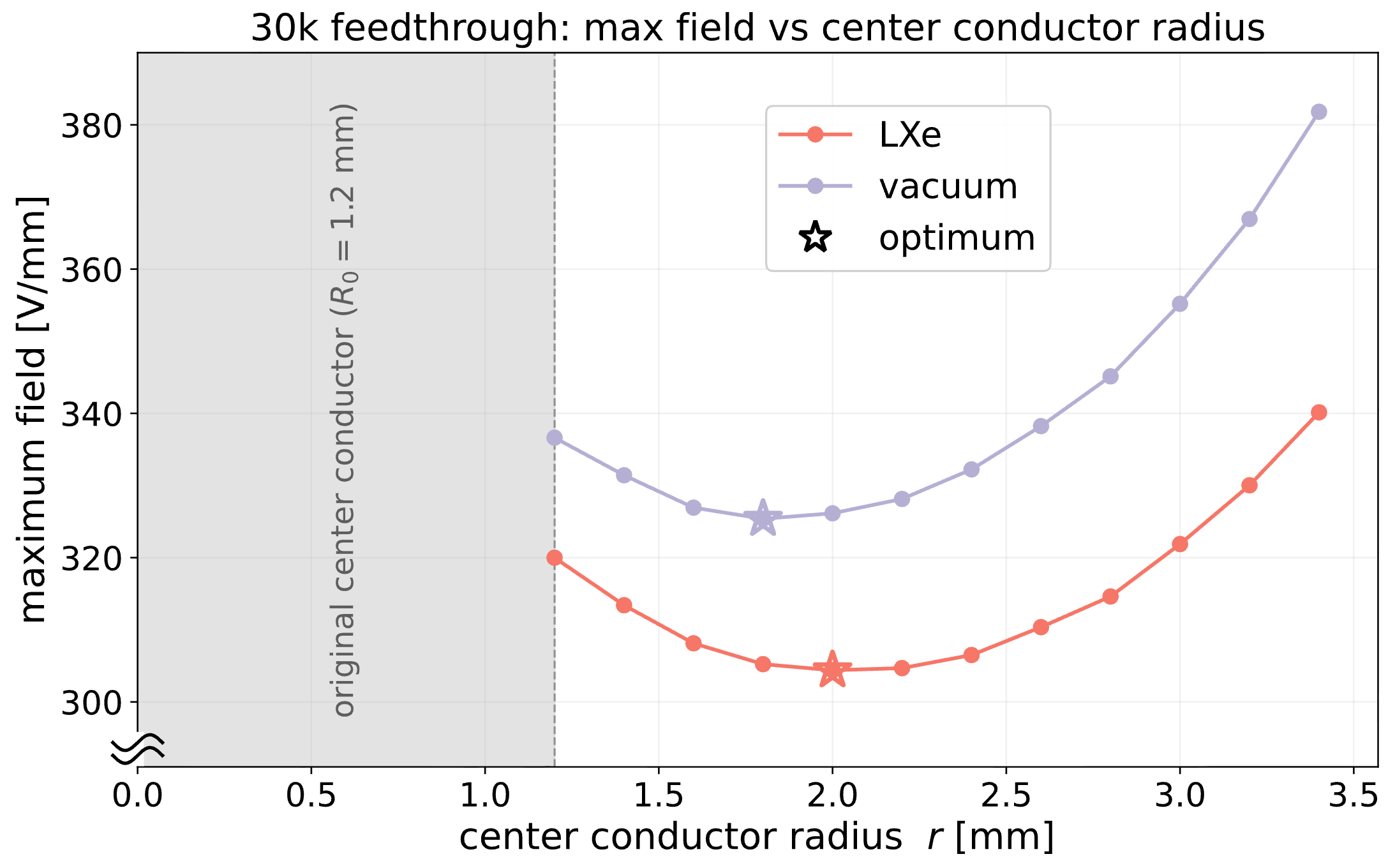}%
\label{fig:scan30k}}
\caption{Maximum electric field as a function of the center conductor radius $r$, in vacuum (purple) and LXe (red). (a)~\SI{100}{\kilo\volt} feedthrough, with the original center conductor radius $r_0 = \SI{2}{\milli\metre}$. (b)~\SI{30}{\kilo\volt} feedthrough, with the original center conductor radius $r_0 = \SI{1.2}{\milli\metre}$. The stars mark the FE-grid optima; the shaded region indicates radii $r < r_0$, which are inaccessible.}
\label{fig:scan}
\end{figure*}

For the \SI{100}{\kilo\volt} feedthrough (\cref{fig:scan}(a)), the peak field decreases steeply from the as-built value to a minimum near $r \simeq \SI{7}{\milli\metre}$ in both vacuum and LXe, and rises again beyond as the conductor approaches the alumina inner radius. The optimum radius from the FE solver lies within \SI{1}{\milli\metre} of the simplified analytical predictions, and the FE peak-field reductions are smaller than the analytical values by about five percentage points, consistent with the additional contribution of the flange geometry.

For the \SI{30}{\kilo\volt} feedthrough ($b = \SI{5.7}{\milli\metre}$, $c = \SI{11.1}{\milli\metre}$, \cref{fig:scan}(b)), the FE-optimum radius is ${\sim}50\text{--}70\%$ larger than the as-built value, but the corresponding field reduction is only a few percent, substantially smaller than the analytical prediction of ${\sim}15\%$. This discrepancy is consistent with the flange geometry playing a relatively larger role at this smaller scale. In this case, the field difference between the vacuum and the LXe cases is also larger. The complete set of values for the four cases is given in \cref{tab:opt}. We stress that increasing the conductor radius does not change the parameters of the alumina insulator, whose thickness is set by the manufacturer; it only reduces the gap filled by the inner medium. The field in the bulk alumina increases accordingly, but at \SI{100}{\kilo\volt} it remains below \SI{1}{\kilo\volt\per\milli\metre}, more than an order of magnitude below the dielectric strength of alumina~\cite{Mieller:2019}.

\begin{table*}[!t]
\caption{Original ($r_0$) and Optimum ($r^{\star}$) Center Conductor Radii, With the Corresponding Peak-Field Reductions, for the Two Commercial Feedthroughs Studied Here\label{tab:opt}}
\centering
\begin{tabular}{lccccccc}
\hline
\textbf{Medium}
    & $r_0$ (mm)
    & $r^{\star}_{\text{an}}$ (mm)
    & $r^{\star}_{\text{FE}}$ (mm)
    & $E_{\text{FE}}(r_0)$ (V/mm)
    & $E_{\text{FE}}(r^{\star})$ (V/mm)
    & $\Delta E_{\text{an}}$
    & $\Delta E_{\text{FE}}$ \\
\hline
\multicolumn{8}{l}{\textit{\SI{100}{\kilo\volt} feedthrough, Kurt J.\ Lesker EFT1C12156A}}\\
Vacuum & 2.0 & 6.4 & 7.0 & 169.2 & 123.9 & $-32.4\%$ & $-26.8\%$ \\
LXe    & 2.0 & 6.9 & 7.0 & 161.0 & 112.6 & $-34.9\%$ & $-30.1\%$ \\
\hline
\multicolumn{8}{l}{\textit{\SI{30}{\kilo\volt} feedthrough, Kurt J.\ Lesker EFT3012093}}\\
Vacuum & 1.2 & 2.2 & 1.8 & 336.6 & 325.4 & $-13.1\%$ & $-3.3\%$  \\
LXe    & 1.2 & 2.4 & 2.0 & 320.0 & 304.4 & $-15.0\%$ & $-4.9\%$  \\
\hline
\multicolumn{8}{p{6.9in}}{\footnotesize Analytical values from \cref{eq:astar}; FE values from the grid scans of \cref{fig:scan}. $\Delta E_{\text{an}}$ and $\Delta E_{\text{FE}}$ are the reductions at the respective optimum radii ($r^{\star}_{\text{an}}$ and $r^{\star}_{\text{FE}}$); $E_{\text{FE}}$ values are at the center conductor with \SI{-1}{\kilo\volt} applied.}
\end{tabular}
\end{table*}

\section{Sleeve Retrofit}
\begin{figure}[!t]
\centering
\includegraphics[width=3.4in]{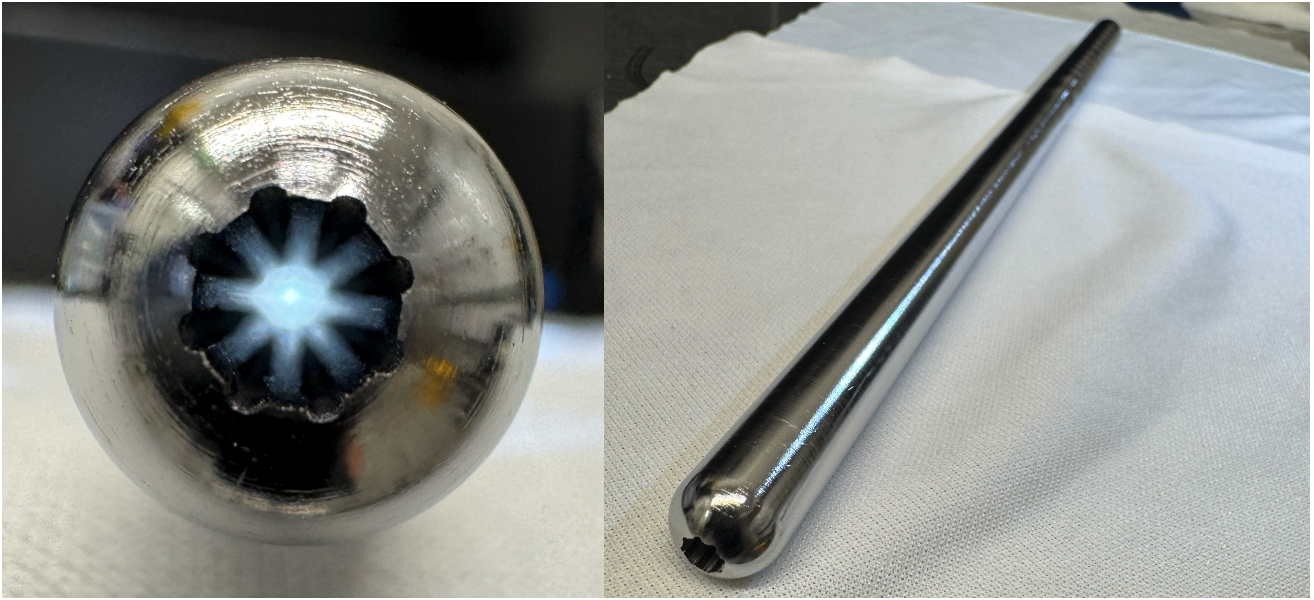}
\caption{Stainless-steel sleeve used to increase the effective radius of the center conductor of the commercial \SI{100}{\kilo\volt} feedthrough. Left: end-on view of the bore matching the original \SI{2}{\milli\metre}-radius pin; the eight axial flutes ensure repeatable centering during assembly and preserve pumping conductance. Right: side view of the sleeve, with hemispherical ends to suppress field enhancement.}
\label{fig:sleeve}
\end{figure}

A metallic sleeve fitted over the existing pin (\cref{fig:sleeve}) provides a practical way to increase the effective radius of the center conductor. In the prototype, built for the \SI{100}{\kilo\volt} feedthrough, the sleeve was machined from 316L stainless steel, with an outer radius of \SI{7.0}{\milli\metre}, an inner radius of \SI{2.0}{\milli\metre} matching the original center conductor, and hemispherical ends. The bore is made slightly oversize relative to the pin, so that the slight intrinsic curvature of the center conductor ensures at least three points of electrical contact along its length. The sleeve has a mass of about \SI{0.38}{\kilo\gram}. With the feedthrough mounted vertically, as in the setup used here, it adds an axial load of only about \SI{4}{\newton} on the conductor, and the free end of the conductor is in any case fixed by its connection to the high-voltage lead, so that no significant bending load is applied to the conductor or to the brazed joints. In other orientations, the free end may have to be supported, for example by a dielectric centering device. 
A concern, for high-purity applications, is to retain the ability of efficiently pumping down all air volumes. This is achieved by a rifling on the inner surface of the sleeve, as illustrated in the figure. While a proper helicoidal rifling would be better at centering the sleeve, the prototype has straight flutes cut by electrodischarge machining,\footnote{Wire EDM performed by Spencer Corporation, Santa Clara, CA, USA.} for convenience. The wire EDM uses a brass wire, which leaves a thin zinc-rich recast layer (${\sim}\SI{2.5}{\micro\metre}$) on the machined surfaces. Because zinc is detrimental to vacuum, this layer was removed by passivation followed by electropolishing. 

\section{Conclusion}
It has been shown that the peak surface field on the center conductor of two examples of commercial ceramic HV feedthroughs is not optimized and can be reduced by up to ${\sim}30\%$ by increasing the radius of the center conductor. The reason why such an optimization is not part of the commercial design is unclear. Possibly larger conductor radii, making the center conductor stiffer, may risk stressing the brazed joint in the event of careless handling. The non-optimal design is presumably justified since those feedthroughs are meant for ultra-high vacuum operation. The analysis shows that, when a feedthrough is used with a different inner medium, optimizing the center conductor radius for that medium can provide substantial gains. A simple mechanical retrofit has also been provided for increasing this radius, without compromising the outgassing properties of the feedthrough. 

\section*{Acknowledgment}
The authors thank R.\ DeVoe and M.\ Vidal for helpful discussions throughout this work and for their comments on the manuscript. They also gratefully acknowledge the support and advice of colleagues in the nEXO Collaboration. The geometry files, mesh scripts, and finite-element input decks used for this study are available from the corresponding author on reasonable request. This work was funded by the US DOE's Office of Science under grant DE-SC0017970.

\bibliographystyle{IEEEtran}
\bibliography{references}

\vfill

\end{document}